# Simultaneous Observation of Solar Neutrons from the ISS and High Mountain Observatories in association with a flare on July 8, 2014

Y. Muraki[1], D. Lopez[1,2], K. Koga[3], F. Kakimoto[4], T. Goka[3], L.X. González[5], S.Masuda[1], Y. Matsubara[1*], H. Matsumoto[3], P. Miranda[2], O. Okudaira[3], T. Obara[3], J. Salinas[2], T. Sako[1], S. Shibata[6], R.Ticona[2], Y. Tsunesada[4], J.F. Valdés-Galicia[7], K. Watanabe[8], T. Yamamoto[9]

1 *Solar-Terrestrial Environment Laboratory, Nagoya University, Nagoya 464, Japan*
2 *Instituto de Investigaciones Fisicas, UMSA, La Paz, Bolivia*
3 *Tsukuba Space Center, JAXA, Tsukuba 305-8505, Japan*
4 *Department of Physics, Tokyo Institute of Technology, Tokyo 152-0033, Japan*
5 *Instituto de Geofisica, Unidad Michoacan, UNAM, Morelia 58089, Mexico*
6 *Engineering Science Laboratory, Chubu University, Kasugai 487-0027, Japan*
7 *Institute de Geofisica, UNAM, 04510, Mexico D. F., Mexico*
8 *Solar-C group, National Astronomy Observatory, Mitaka 181-8588, Japan*
9 *Department of Physics, Konan University, Kobe 658-8501, Japan*
E-mail: `muraki@stelab.nagoya-u.ac.jp`

**Abstract**

An M6.5-class flare was observed at N12E56 of the solar surface at 16:06 UT on July 8, 2014. In association with this flare, solar neutron detectors located on two high mountains, Mt. Sierra Negra and Chacaltaya and at the space station observed enhancements in the neutral channel. The authors analysed these data and a possible scenario of enhancements produced by high-energy protons and neutrons is proposed, using the data from continuous observation of a solar surface by the ultraviolet telescope onboard the Solar Dynamical Observatory (SDO).

1. Introduction

On July 8, 2014, a solar flare occurred on the solar surface at N12E56. According to the GOES observation, the flare started at 16:06 UT and peaked at 16:20 UT, with a flare intensity of M6.5. In association with this flare, hard X-rays were detected by the GBM instrument onboard the FERMI satellite. The intensity of the hard X-rays peaked at 16:13 at around 400 photons/detector/sec within the range 100-300 keV. The solar flare was continuously monitored before the flare by the ultraviolet telescope onboard the Solar Dynamical Observatory (SDO).

We have searched for a solar neutron signal associated with this flare in the data registered by the solar neutron detectors located at observatories in Mt. Chacaltaya, Bolivia and Mt. Sierra Negra, Mexico. Possible neutron signals were also detected by the solar neutron detector SEDA-NEM onboard the International Space Station (ISS). This paper describes the data observed in these detectors. A possible scenario of the neutron production mechanism at the Sun will be provided.

## 2. Detectors

### 2.1 Solar Neutron Detector (SND) at Mt. Chacaltaya

The Chacaltaya observatory is located at an altitude of 5,250 m. The geographical location is S16.2°,W68.1°, nearly 30km from La Paz, Bolivia. The vertical atmospheric depth at the observatory is almost half that of sea level ($540 g/cm^2$). This is a great advantage in observing cosmic rays and solar energetic particles (SEP) because of smaller atmospheric attenuation compared to cases in ground-based experiments. Accordingly, solar neutrons are expected to arrive there with less absorption in the atmosphere.

In September 1992, a solar neutron detector was added in the observatory. The solar neutron detector (SND) comprises four plastic scintillation counters, each of which has an area of $1m^2$ and within which 40 cm-thick plastic scintillators are installed. The solar detector is covered by anti-counters made of plastic scintillator, which allows to distinguish neutral and charged particles. Furthermore the deposited energy in the plastic scintillator are measured with four-level discriminators, namely >40, >80, >160 and >240 MeV. The discriminator level is calibrated using the energy deposited by muons traversing the plastic scintillator. (When muons penetrate the 40 cm thick plastic scintillator vertically, they leave about 80 MeV in the scintillator by ionization loss.) The typical trigger rate of SND is about 55,000 (per $4m^2$/min at >100 MeV). The details of Chacaltaya SND have been provided in elsewhere [1].

### 2.2 Solar Neutron Telescope (SNT) at Mt. Sierra Negra

The solar neutron telescope (SNT) of Mt. Sierra Negra is located at (97.3°W,19.0°N), about 200 km east from Mexico City DF. At the altitude 4,580 m the SNT is set. The average vertical atmospheric pressure is $575 g/cm^2$. The SNT has been operating on the site since June 2003.

The detector comprises a $4m^2$ plastic scintillator of 30 cm thick with a detection area equivalent to the Chacaltaya SND, however the plastic scintillator is 10 cm slimmer than that of the SND. The anti-counter at Sierra Negra SNT is made of proportional counters. On top of the detector, 5 mm-thick lead plates are installed to increase the detection efficiency for gamma-rays. The trigger rate of SNT within one minute is about 8,000 counts per detector (per $4m^2$/min) for the neutral particles with energy exceeding >100 MeV. The SNT includes a set of proportional counters to determine the arrival direction of entrance particles, which is why it is called a Solar Neutron Telescope (SNT). Technical details of the SNT are available in [2-5].

### 2.3 Solar Neutron Detector Onborad ISS

Solar neutron detection in space, we will have a merit that we are free from the absorption in the atmosphere. The solar neutron detector in space (NEM) was prepared as one of the detectors of the the Space Environment Data Acquisition System (SEDA). The actual launch of SEDA to the ISS was conducted on July 16, 2009 by the space shuttle Endeavor, since then the SEDA-NEM detector has been continuously monitoring neutrons from the Sun. Details of the SEDA-NEM detector can be found in [6,7]

The background neutrons are mainly induced by cosmic rays with interaction between the ISS materials. Such neutrons can be identified from their arrival directions and the background can be reduced by a factor ~1/36, when we look the Sun with the solid angle of $2/3\pi$ steradians. The detection efficiency of the sensor was estimated by GEANT4 package at around ~2 percents within the energy range between 40-150 MeV [7].

## 3. Observed Data and Analysis

### 3.1 Time profiles of Event Observed by Chacaltaya Detector

The 2-minute counting rates observed by the Chacaltaya SND are presented in the upper panels of Figure 1. The data are taken by the SND around the flare time 15:51-17:15 UT on July 8, 2014 in the four different neutral threshold channels, >80MeV and >160 MeV. (Due to the limit of the paper, we have plotted only two channels.) The local time 11:51-13:15 corresponds to the local noon at Chacaltaya observatory and the Sun was near the zenith when the flare occurred around 12:06 LT. The total atmospheric pressure to the solar direction was 623 g/cm$^2$.

Since a good linear correlation is known between the data of the neutral channel and the charged channels, the variation of the charged channel is used to estimate the background for the neutral channel. We examined the correlations between two channels using the same period of a different day when no strong solar flare was reported, July 6, 2014. The correlation coefficients for 20-minute running averages between charged and neutral channels are found as 0.92, 0.97, 0.98 and 0.97 for the four energy ranges; >40, >80, >160, and >240 MeV respectively. This implies that the counting rate of the charged channel can be used as a measure for the background level of the neutral channel above which we find neutron signals.

In order to investigate whether the counting rate follows the Gaussian shape distribution or not, we have examined the data of the charged channel and the neutral channel with the threshold energy >160 MeV recorded in the time during 15:31 and 19:07 UT. The deviations of the 2-minute counting rate from the 10-minute running average data of the charged channel and neutral channel can be expressed by the Gaussian distribution with the standard deviation of σ=233 and 211 respectively. Therefore we will be able to estimate the statistical significances that were observed in the neutral channels as found in the upper panel of Figure 1. The statistical significances of each excess are listed in Table I. When we draw the background line on the data of the neutral channels in Figure 1, the counting rate of charged channel was scaled down to meet with counting rate of neutral channel for each energy range; actually the numbers 2.19, 2.12, 2.2, and 2.51 are applied for each channel of >40, >80, >160 and >240 MeV respectively.

The integral energy distributions of neutrons detected with the Chacaltaya SND are presented in Figure 2. The shape of each distribution is in reasonable agreement between the three excesses at different times after the flare. The data of a different solar flare observed on September 7, 2005 is also shown in Figure 2 for comparison [8-12]. The data suggest that the intensities of neutrons emitted in the 2014 event were smaller than that of the 2005 event, however the integral power indexes are slightly harder than the event of September 7, 2005 with the power index of γ= -2.0 [8].

### 3.2 Time profiles of Event Observed by Sierra Negra Detector

Here we present the time profile of the SNT at Sierra Negra during the flare observation period 15:50-17:16 UT. The lower panels of Figure 1 show the 2-minute value of the neutral channels exceeding >90 MeV and >120 MeV respectively. The dotted line with error bars show the estimated background line from the charged channels. The background is estimated by the 10-minute running average of the charged channel. The statistical significances of 2-minute and 10-minutes values of each time are shown in Table II.

It is interesting to compare the data of >120 MeV at Sierra Negra with that of >40 MeV, >80 MeV and >160 MeV at Chacaltaya SND. The first peak of Sierra Negra at 16:07 UT coincides well with the first peak of Chacaltaya enhancement at 16:07 UT. The second peak at around 16:32-16:40 UT also coincides with the second peak observed at Chacaltaya at 16:41 UT in the channels of >40 MeV, >80 MeV, and >160 MeV. Accordingly, we consider these enhancements to be inter-related and produced by solar neutrons entering at the top of the atmosphere.

The effective atmospheric depth to solar neutrons during the flare time at Sierra Negra was 617 g/cm$^2$. In the estimation of the atmospheric depth, the large angle quasi-elastic scattering effect between neutrons and atmospheric nuclei is taken account [13-14].

### 3.3 Energy Spectrum at Top of Atmosphere

Here we try to reduce the neutron intensity at the top of the atmosphere. In this process, we convert the observed counting rate by three corrections factors. The numbers are given in Table III. First we must correct the attenuation loss. The attenuation of neutrons in the atmosphere has been evaluated by [15,16]. Then we correct the data by the detection efficiency of the SND and SNT. The detection efficiency has been evaluated by two of present authors independently based on the MC calculations [10-12] and by the accelerator experiment [17].

Recently, the relation between the threshold energy and the average incident energy has been estimated by the Monte Carlo calculation using GEANT4 package [16]. The results show that for threshold energies of >40, >80, >160 and >240 MeV at Chacaltaya altitude, the average incident energy is expected to be 100, 250, 1,000 and 2,500 MeV respectively for incident neutrons entering with an angle to the atmosphere of between 0 (vertical) and 30 degrees [18]. The data points presented in Figure 3 are prepared based on these values. The correction of the neutron decay has not yet been included in Figure 3. It is impressive to know that the energy spectrum of September 7, 2005 event can be expressed by the power law with the integral index $\gamma=-2.0$ as predicted by [8].

### 3.4 Observation of Event by SEDA-NEM in Space

During the GOES start time of the flare (16:06 UT) to the GOES maximum time (16:20 UT), the ISS was flying over the night region of the Earth. When it passed over the central Pacific region on 16:21 UT, the ISS entered into the region where the solar flares could be observed. During the above period, we missed a chance to observe a possible high-energy neutrons emitted at ~16:07 UT that was detected by the solar neutron detectors at Mt. Chacaltaya and Sierra Negra. However we had a chance to see low-energy neutrons with the energy range of 35 - 70 MeV. Because solar neutrons with an energy of 70 MeV are expected to arrive the Earth 14 minutes later than light. Therefore we have searched solar neutrons in the data of SEDA-NEM carefully. Consequently, we found 30 candidates of solar neutron events in the data. They were detected in the time 16:34 UT-16:48 UT. The selection of events were made, applying of the directional discrimination function to the background. Total amount of the counting rate of the SEDA-NEM during 16:34-16:48 UT was 226 events.

Since SEDA-NEM detector measures the energy of ecah neutron, the fight time of neutrons in the interplanetary space can be calculated. The start time distribution at the Sun indicates two origins. Solar neutrons seem to be produced at least twice in this flare, aound ~16:16 UT and ~16:32 UT. We present the energy spectrum for each production in Figure 5 under this assumption.

### 3.5 Energy Spectrum at Sun

In order to convert the spectrum at the top of the atmosphere into the spectrum at the Sun, we take account of the flight time difference of each neutron and decay factor. As we know, neutrons with the energy of 100 MeV will arrive at the Earth 11 minutes later than light. Therefore we have combined different time counting rates observed in the different channels and produced an energy spectrum. This procedure corresponds that *the clock of neutrons is synchronized at the Sun*. The results of the MC calculation on the most probable energy for each threshold energy are also applied [16,18]. The results are shown in Figure 4. The mark (●) corresponds to the flare observed on September 7, 2005, while the other marks (Δ) present the data observed on July 8, 2014 by the SND at 16:07 UT. The data point indicated by the mark (♠) is obtained by the SEDA observation under an assumption that those neutrons were produced instantaneously around ~16:16 UT.

From Figure 4, the integral fluxes at >50 MeV and >1,000 MeV are predicted as to be $5.15 \times 10^5$/ (m²•min) and $1.57 \times 10^4$/ (m²•min) respectively. Then the neutron flux at the Sun is estimated as to be $(1.5\pm0.3) \times 10^{29}$ neutrons/steradians at >50 MeV and $(4.5\pm0.3) \times 10^{27}$ neutrons/steradians at >1000 MeV respectively.

4. Possible Interpretaion on Production Mechanism of Solar Nuetrons

According to our Monte Carlo simulation, the excess at 16:07 UT of the neutral channel of SDO at Chacaltaya with a discriminator level of >160 MeV may be induced by neutrons with incident energy of ~1,000 MeV. Therefore, we carefully investigated two pictures of the SDO, particularly photos taken at around 16:05 UT (Figure 6). Then we found the collision between two loops indicated by teh whire arrow in Figure 6. The collision forms a x-type crossing structure between loops. As the two loops intersect, a strong electric field may be formed, which may cause protons to accelerate up to ~10 GeV [19]. When these protons fall down and collide with the solar atmosphere, neutrons with energies of ~1,000 MeV are generated instantaneously. The first peak of the Chacaltaya SDO and Sierra Negra SNT may possibly be produced by this mechanism [20,21].

Then, how to best interpret the later enhancement observed by the SND at Chacaltaya that continued for more than 26 minutes? The excess observed by the channel >160 MeV continued from 16:35 to 17:01 UT. Neutrons observed by >160 MeV correspond to a primary energy of 1,000 MeV and they were delayed only ~1 minute compared to the light. We cannot explain the 26-minute long enhancement observed at the channel of >160 MeV with the impulsive production model. Accordingly, this enhancement must be attributable to a continuous production mechanism of high-energy protons at the Sun. The later peak started at ~16:34 UT. Taking the 1-minute delay into account, the production of those neutrons must continue at the solar surface between 16:33 and 17:00 UT.

Therefore we carefully re-investigated the SDO pictures, but only recognized two bright parallel ribbon flares suspended in the north-south direction at the solar surface. According to FERMI-GBM observation, while the intensity of hard X-rays was strong between 16:12 and 16:27 UT, it did not continue beyond 16:27 UT. It is therefore quite hard to explain the later peak of SND by the same acceleration model applied to the first peak.

What we would like to introduce here is a hypothesis stating that these high-energy particles are produced at the shock front of the CME leveraging the perpendicular shock acceleration process [22]. However, we cannot see the acceleration site because it may be far from the solar limb and detecting such region may be very difficult, even using the excellent onboard detector. Once, Tsuneta and Naito proposed a perpendicular shock acceleration model to the downside direction on the solar surface [23], which can also be applied in an upside direction. Via this acceleration mechanism, particles may be accelerated at the front shock of the CME, trapped in the CME magnetic loop and fall down to the solar surface, where they produce high-energy neutrons. Protons could be accelerated beyond ~10 GeV and the seed particles of such high-energy protons may be already prepared between 16:12 UT and 16:17 UT when the two loops collide over the active region (as presented in Figure 7). Within this plausible scenario those particles were already injected in the CME magnetic loop at ~16:17 UT and were accelerated beyond 10 GeV by the perpendicular shock acceleration mechanism. It took about ~17 minutes for the acceleration time into high energies. These accelerated protons fell down continuously on the solar surface [24] through the magnetic loop and might produce neutrons after 16:34 UT.

Once Cheng et al pointed out that the collision between the magnetic ropes used to happen in the following way: via the collision between the pre-existing flux rope and the rising hot magnetic rope, a dynamical CME is formed [25]. Based on the hypothesis pointed out by Cheng et al., protons may be accelerated into high energies around the very turbulent region at the top of the magnetic rope by the perpendicular shock acceleration mechanism. The dynamical picture Figure 7 supports this point of view.

5. Conclusion

In association with the M6.5 flare on July 8, 2014, the solar neutron detectors located at Chacaltaya, Sierra Negra and International Space Station observed several enhancements over the background. The enhancement observed in the neutral channel can be naturally explained if we consider them attributable to solar neutrons.

The production mechanism cannot be explained by a single model, but at least one of the enhancements may be explained by the electric field generated by the collision of loops and the other by the shock acceleration mechanism at the front side of the CME. However, we cannot identify the acceleration site using current detectors although observations help us to determine where and when it happened.

The excess of SND at 16:07 UT was produced by particles generated in small-sized magnetic loops (shown by Figure 6). The long continuous enhancement observed by SND must be produced by completely different mechanisms from those pictures above, which must be related with the CME (Figure 7), whereby the particles concerned precipitate over the solar surface and produce high-energy neutrons [26]. Those neutrons are then registered by detectors located at Chacaltaya and Sierra Negra. The Sun is a particle accelerator and most probably has the ability to accelerate ions beyond ~10 GeV.

**Table III**

| Threshold En. | incident energy | attenuation | efficiency | survival factor |
|---|---|---|---|---|
| > 40 (MeV) | 100 ± 20 (MeV) | 0.0003 ±0.0002 | 0.17 ± 0.02 | 0.38 |
| > 80 | 250 ± 70 | 0.01 ±0.003 | 0.12 ± 0.02 | 0.56 |
| >160 | 1000 ± 250 | 0.05 ±0.01 | 0.12 ± 0.02 | 0.8 |
| > 240 | 2500 ± 300 | 0.16 ±0.02 | 0.10± 0.02 | 0.9 |

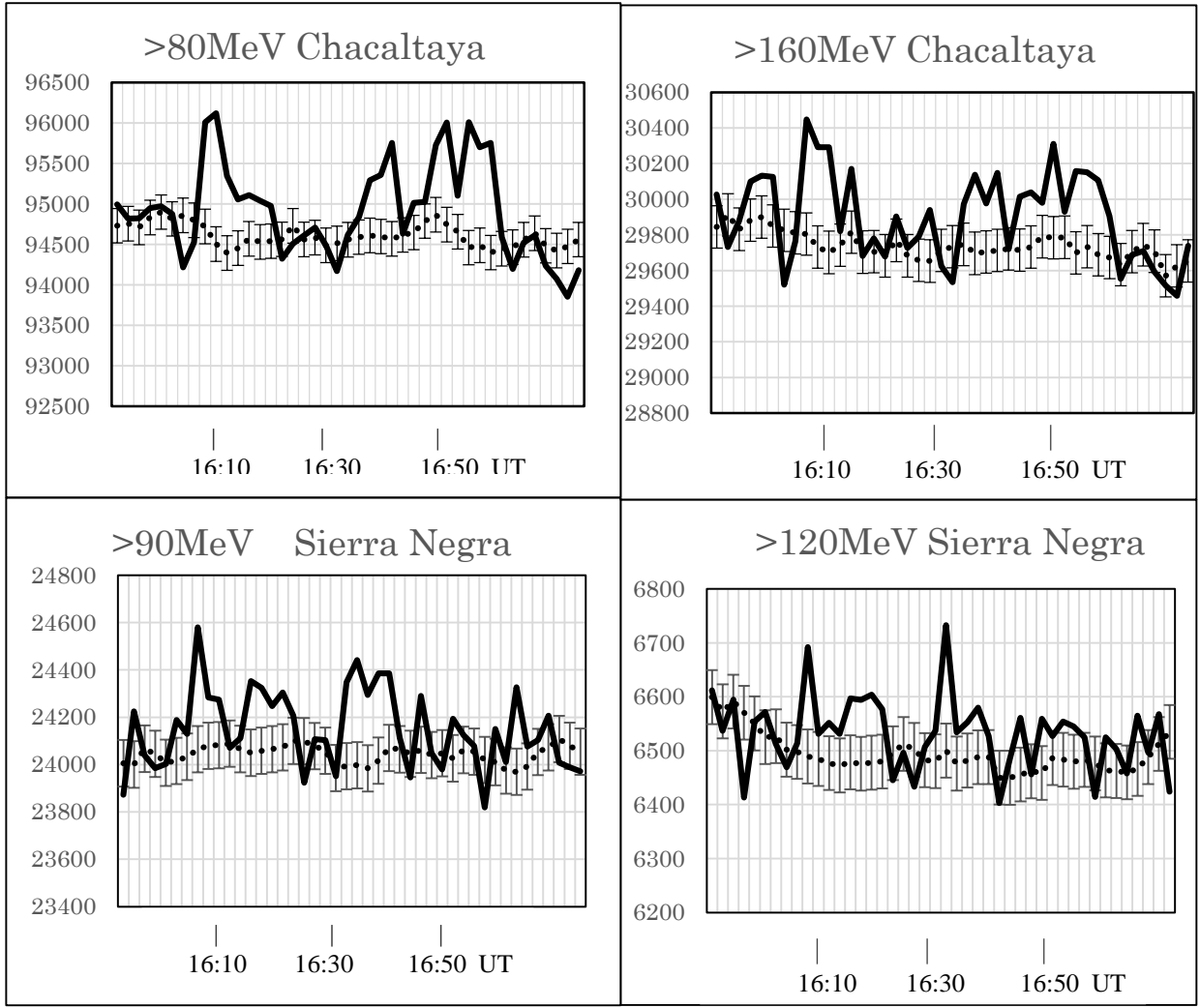

**Figure 1.** Time profile of event observed at Chacaltaya (upper panel) and Sierra Negra (lower panel).

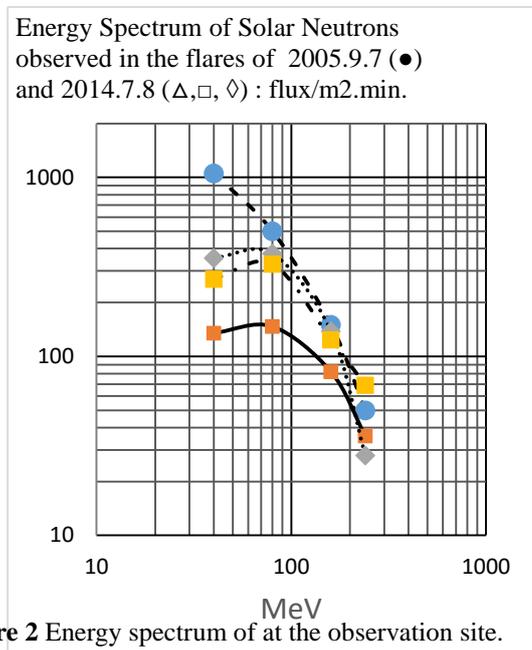

**Figure 2** Energy spectrum of at the observation site.

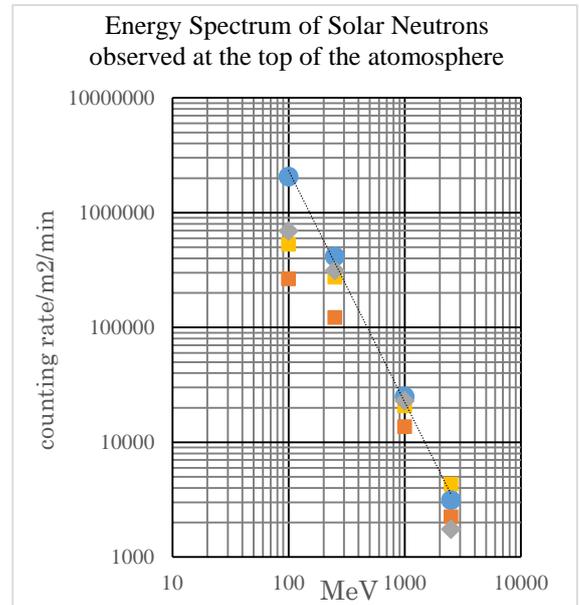

**Figure 3** Spectrum at the top of atmosphere

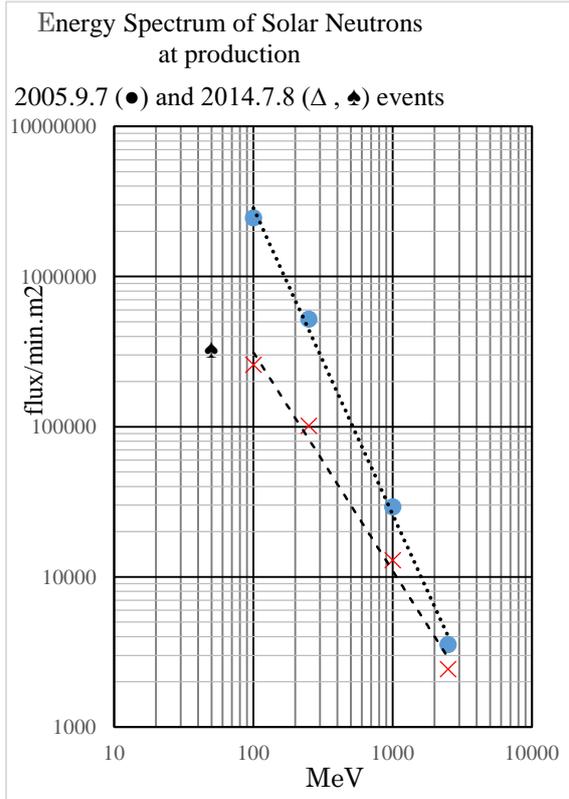

**Figure 4.** The energy spectrum of solar neutrons at the production site (at the Sun).

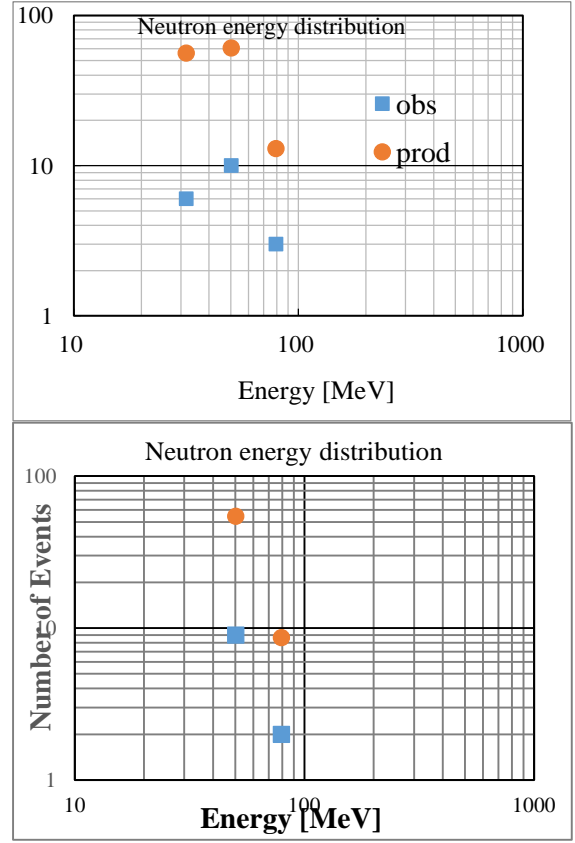

**Figure 5.** Neutron energy distribution by SEDA

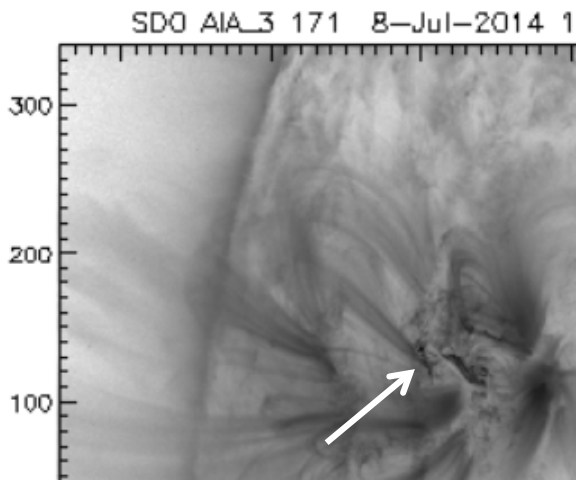

**Figure 6.** SDO 171nm data on 16:04:59 UT.

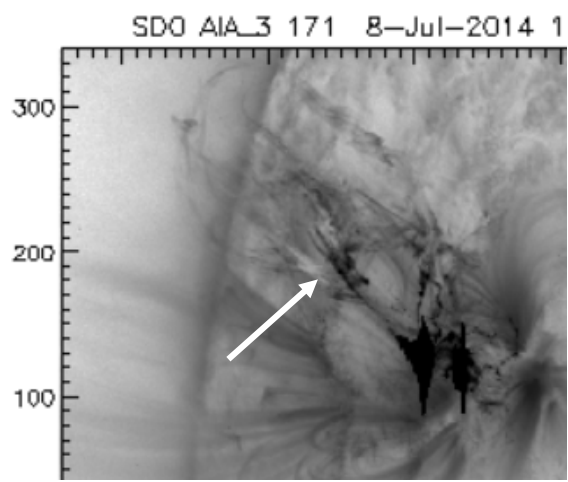

**Figure 7.** SDO data on 16:15:35 UT.

**Table I** Chacaltaya statistical significance

Duration | >40MeV  >80MeV  >160MeV  >240MeV | 12NM64

2014.7.8 UT (in minutes)      all in σ

   16:07-16:13 (6) | 6.2  7.2  6.0  4.9 | 5.2

   16:37-16:43 (6) | 2.4  4.9  3.8  1.6 | 2.9

   16:45-16:57(12)| 4.9  8.1  5.8  5.5 | 8.7

2005.9.7 UT (in minutes)

   17:34-17:38 (4) | 12  8.8  5.5  3.2 | 15

**Table II** Sierra Negra statistical significance

    >90 MeV   >120 MeV

// 2014.7.8  16:02-16:04 (2) | 5.3  0(no excess)

          16:06-16:08 (2) | 2.0  4.1

          16:12-16:22 (10)| 3.2  2.8

          16:32-16:34 (2) | 4.5  4.7

          16:32-16:40 (8) | 4.6  2.8